\documentclass[twocolumn]{aastex63}
\usepackage{xspace,amsmath,amssymb,lineno,multirow}

\newcommand{\GG}[1]{}

\shorttitle{X-ray Sources in the SSA22 Chandra Field}
\shortauthors{Radzom et al.}

\graphicspath{{./}}

\begin{document}

\title{X-ray Sources in the SSA22 Chandra Field}

\correspondingauthor{Brandon T. Radzom}
\email{bradzom@iu.edu}

\author[0000-0002-0015-382X]{Brandon T. Radzom}
\affiliation{Department of Astronomy, Indiana University,
727 East 3rd Street, Bloomington, IN 47405-7105, USA}
\affiliation{Department of Astronomy, University of Wisconsin-Madison,
475 N. Charter Street, Madison, WI 53706, USA}

\author[0000-0003-1282-7454]{A.~J.~Taylor}
\affiliation{Department of Astronomy, University of Wisconsin-Madison,
475 N. Charter Street, Madison, WI 53706, USA}

\author[0000-0002-3306-1606]{A.~J.~Barger}
\affiliation{Department of Astronomy, University of Wisconsin-Madison,
475 N. Charter Street, Madison, WI 53706, USA}
\affiliation{Department of Physics and Astronomy, University of Hawaii,
2505 Correa Road, Honolulu, HI 96822, USA}
\affiliation{Institute for Astronomy, University of Hawaii, 2680 Woodlawn Drive,
Honolulu, HI 96822, USA}

\author[0000-0002-6319-1575]{L.~L.~Cowie}
\affiliation{Institute for Astronomy, University of Hawaii,
2680 Woodlawn Drive, Honolulu, HI 96822, USA}

\begin{abstract}

The Hawaii Survey Field SSA22 is the fourth deepest Chandra X-ray field. To allow for the fullest exploration of this field, we present new optical spectroscopy from Keck/DEIMOS and Keck/LRIS, which, in combination with the literature, brings the spectroscopic completeness of the 2--8~keV sample to 62\%. We also make optical spectral classifications and estimate photometric redshifts for the sources without spectroscopic redshifts. We then determine hard X-ray luminosity functions (XLFs) for the full sample of Active Galactic Nuclei (AGNs), as well as for the broad-line AGNs (BLAGNs) and the non-BLAGNs separately. Our XLF for the full sample is in good agreement with the literature, showing relatively strong evolution over the redshift range $0.25\le z < 4$. The XLFs for the BLAGNs and the non-BLAGNs imply distinct evolution with redshift, with BLAGNs becoming increasingly dominant at higher redshifts and X-ray luminosities.

\end{abstract}

\keywords{AGN --- X-ray --- quasars --- galaxy evolution}

%%%%%%%%%%%%%%%%%%%%%%%%%%
\section{Introduction} \label{sec:intro}
%%%%%%%%%%%%%%%%%%%%%%%%%%
It is well-established that supermassive ($\geq10^6$~M$_{\odot}$) black holes (SMBHs) reside at the centers of nearly all galaxies (e.g., \citealt{KormendyHo2013} and references therein) and that many of these objects actively accrete gas to become AGNs.
However, observations of AGNs are complicated by the fact that much of their emitted light can be absorbed by the surrounding gas \citep[e.g.,][]{Almaini99,Hasinger2001}, which means optical campaigns that detect AGNs are biased toward objects that exhibit broad emission lines, such as quasars or other unobscured AGNs.

Hard (2--8~keV) X-ray surveys performed with the Chandra X-ray Observatory Advanced CCD Imaging Spectrometer \citep[ACIS;][]{garmire2003} have proven to be an efficient and unbiased probe of AGNs \citep[see, e.g., reviews by][]{Mushotzky2004,BrandtRev2015,HickoxRev2018}, as hard X-rays  are able to pass through high column densities of hydrogen gas ($N_H\sim 10^{23} \text{ cm}^{-2}$) without being scattered or absorbed. Such campaigns have been successful in identifying obscured AGNs, revealing that many SMBHs have been in active stages much more recently than was previously thought. However, it should be noted that even 2--8~keV selection techniques under-count Compton-thick AGNs with $N_H\gtrsim10^{24} \text{ cm}^{-2}$.

The X-ray luminosity function (XLF) is one of the best tools available to characterize the cosmic evolution of AGN accretion activity. 
Early XLF determinations revealed the AGN population to be strongly evolving, particularly at the bright end \citep[e.g.,][]{Cowie2003, Fiore2003, Ueda2003}, with the most luminous AGNs peaking in number density near $z\sim2$, and their fainter counterparts peaking more recently \citep[e.g.,][]{BargerCowie2005,Hasinger2005}. This phenomenon is normally referred to as AGN downsizing. 

The evolution of AGNs is coupled to their spectral type. Unabsorbed or broad-line AGNs dominate the AGN space density at high X-ray luminosities near $z\sim 2$, with their fraction increasing out to $z\sim 3$ \citep{Buchner2015, Georgakakis2017}. In contrast, absorbed or non-BLAGNs largely constitute the faint end of the XLF at lower redshift \citep[e.g.,][]{Yencho2009, Aird2015}, as well as the majority of the AGN luminosity and number density overall \citep[e.g.,][]{Buchner2015}, even after accounting for the effects of torus orientation and host galaxy properties \citep[e.g., dust obscuration;][]{Sazonov2015,Vito2018}.

In this work, we investigate the difference in luminosity evolution for these two classes of AGNs. We present new spectroscopic data for X-ray sources in the Hawaii Survey Field SSA22, one of the deepest and most well-studied of the Chandra deep fields, largely due to the presence of the $z=3.09$ galactic protocluster discovered by \cite{Steidel1998}. Our primary goals are as follows:
\begin{enumerate}

\item To update the SSA22 Chandra point-source catalog redshifts originally published in \cite{Lehmer2009} (hereafter, L09) using our newly obtained spectroscopic redshifts (hereafter, speczs) and spectral classifications, together with those from the literature.

\item To utilize these data to investigate the X-ray luminosity evolution of AGNs and cosmic variance more broadly by constructing XLFs in several intervals over the redshift range $0.25\leq z<4$.

\item To characterize the differential X-ray luminosity evolution of BLAGNs and non-BLAGNs by constructing the XLFs for both populations separately.

\end{enumerate}
We assume the following flat cosmology throughout this work: $H_0=70$ km s$^{-1}$ Mpc$^{-1}$, $\Omega_M=0.3$, $\Omega_\Lambda=0.7$ and $\Omega_R=0$.

%%%%%%%%%%%%%%%%%%%%%%%%%%
\section{Observations and Methods} \label{sec:obs}
%%%%%%%%%%%%%%%%%%%%%%%%%%

%%%%%%%%%%%%%%%%%%%%%%%%%%
\subsection{X-ray Imaging Data and Catalog}
%%%%%%%%%%%%%%%%%%%%%%%%%%

Our full sample consists of 297 X-ray sources in a deep 400~ks Chandra/ACIS-I %(Advanced CCD Imaging Spectrometer) 
observation of the SSA22 field (P.I. D.~Alexander; L09). The field, which was previously observed with Chandra/ACIS-S \citep[P.I. G. Garmire; ][]{Cowie2003}, was centered on the $z=3.09$ galactic protocluster described by \citet{Steidel1998}. The X-ray campaign consisted of four separate observations, ranging from about 70 to 120~ks, taken between October 1 and December 30 of 2007. The survey was conducted over $\sim330$~arcmin$^2$. Sensitivity limits of $\sim 5.7 \times 10^{-17}$ and $\sim 3.0 \times 10^{-16}$ erg s$^{-1}$ cm$^{-2}$ were reached for the 0.5--2~keV and 2--8~keV bands, respectively. 

L09 published the main Chandra catalog (their Table~2), giving each of the 297 sources observed in X-rays a unique source identification number in order of increasing right ascension, which we retain throughout this work.

%%%%%%%%%%%%%%%%%%%%%%%%%%
\subsection{Optical Imaging Data}
%%%%%%%%%%%%%%%%%%%%%%%%%%

L09 matched their X-ray catalog to available near-infrared and optical counterparts. This resulted in up to 10 filter magnitudes for each object, including $K$-band from the United Kingdom Infrared Telescope (UKIRT) Infrared Deep Sky Survey (UKIDSS; \citealt{Lawrence2007}), Spitzer/IRAC \citep{Fazio2004} 3.6, 4.5, 5.8, and 8.0~$\mu$m bands \citep{Webb2009}, and Subaru/Suprime-Cam \citep{Miyazaki2002}  $B$, $V$, $R$, $i'$, and $z'$ bands \citep{Hayashino_2004}.

We performed our own counterpart matching using a new optical catalog constructed from re-reduced imaging from Subaru/Hyper Suprime-Cam \citep[HSC;][]{Miyazaki2018} in the $g$, $r$, $i$, $z$, $y$, NB816, NB921, and NB926 bands (A.~Taylor et al. 2022, in prep). We supplement these data with archival $U$ band data from Canada-France-Hawaii Telescope (CFHT)/MegaCam MegaPipe reductions (observation ID: G012.334.588+00.283), $K$-band data cutouts from UKIDSS-DR11PLUS, and Spitzer/IRAC 3.6, 4.5, 5.8 and 8.0~$\mu$m archival Super Mosaics. 

For each X-ray position, we searched for counterparts in the Subaru/HSC optical catalog using a 1\farcs5 radius. We found 252 optical counterparts for the X-ray catalog. We measured $3''$ diameter aperture magnitudes in the $U$, $g$, $r$, $i$, $z$, $y$, and $K$ bands, and $6''$ diameter aperture magnitudes in the Spitzer/IRAC bands, all centered on the optical counterpart positions. We then performed $3''$--$4''$ ($6''$--$8''$) aperture corrections for each filter to produce our final photometric catalog.

%%%%%%%%%%%%%%%%%%%%%%%%%%
\subsection{Optical Spectroscopy}
%%%%%%%%%%%%%%%%%%%%%%%%%%
\label{sec:speczs}
Redshift identifications for the X-ray sources have previously been reported by L09 and \cite{Saez2015}. We conducted spectroscopic observations to improve the completeness of both the speczs and the spectral classifications for the X-ray sources.

We obtained speczs from our own spectra for 147 sources. 
These spectra include 126 from the Deep Imaging Multi-Object Spectrograph (DEIMOS; \citealt{Faber_2003}) on Keck~II, and 50 from the Low-Resolution Imaging Spectrometer (LRIS; \citealt{Oke_1995}) on Keck~I. Only 21 of the LRIS spectra are unique. 
For both instruments, we took three 20~min sub-exposures on each slitmask, 
dithering $\pm 1\farcs{}5$ along the slits for improved sky subtraction and
minimization of CCD systematics. Each source received a total exposure time of 1~hr.

Comparing our speczs with the literature, we determined that one is likely not the correct counterpart to the X-ray source but rather a foreground contaminant (source ID 142; see Figure 2 in \citealt{Umehata_2019}), while another (a star) is likely a misidentification (source ID 62; \citealt{Saez2015}). Thus, our spectra update the speczs for a total of 145 sources in the field (including stars), 41 of which did not have a previously measured specz.

In constructing the final updated catalog, we adopt six speczs from L09 and 23 from \cite{Saez2015} (including source ID 62 mentioned above). 
We further adopt six speczs from \cite{Umehata_2019} (including source ID 142 mentioned above), one from \cite{Lilly1995}, one from \cite{Newman2010A}, and one from SDSS~DR13 \citep{Albareti_2017}. We note that several of our new redshift measurements exactly match those previously reported in the literature, including nine from \cite{Saez2015}, five from L09, and one from \cite{Mawatari_2021}. In total, we have speczs for 183 X-ray sources, 17 of which are stars.
This corresponds to a spectroscopic completeness of 62\%, close to the maximum completeness observed in any of the deep X-ray fields.

\begin{figure}[h!]
\centering
\includegraphics[width=\columnwidth]{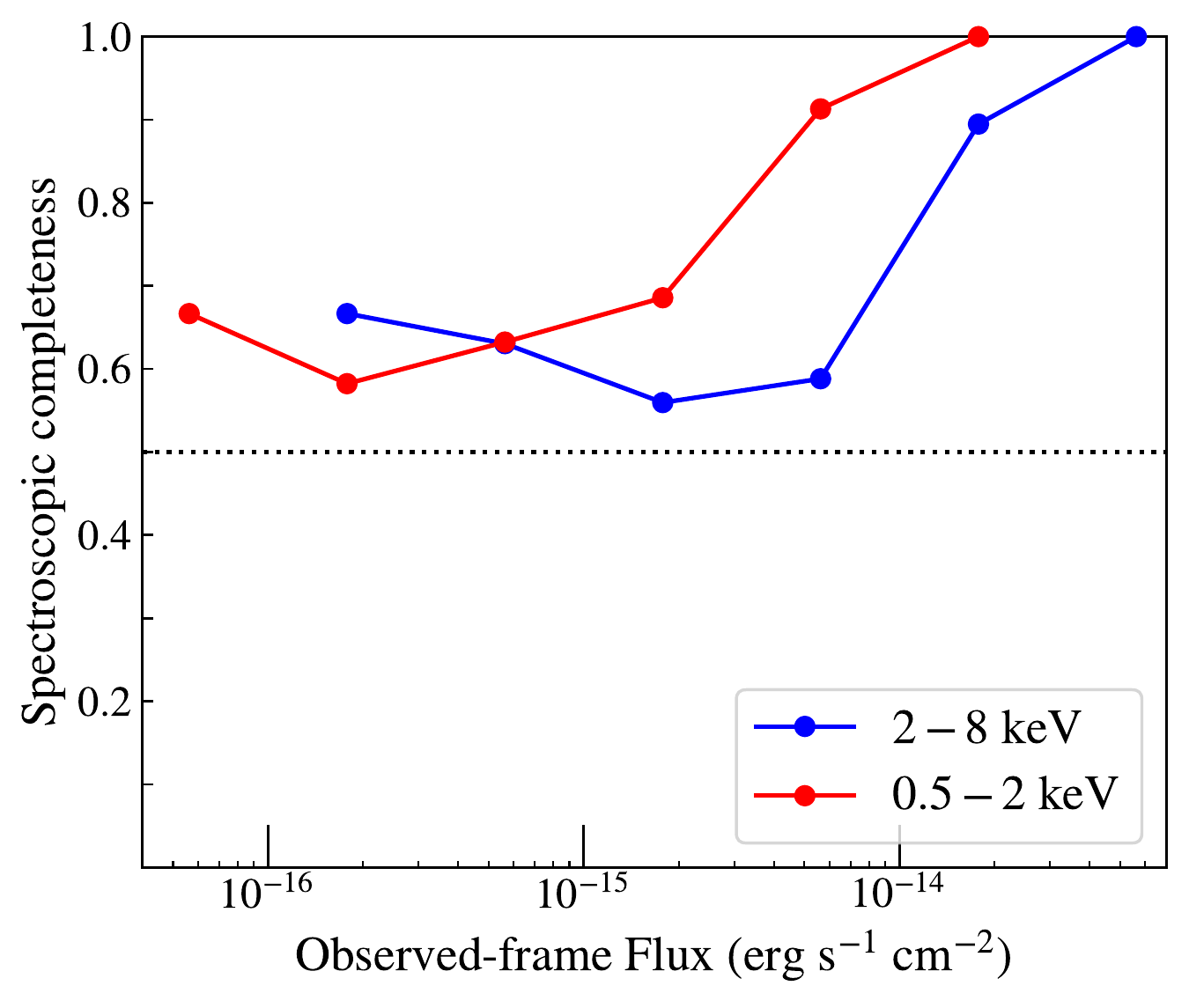}
\caption{Fraction of spectroscopically identified sources
as a function of the observed-frame 2--8~keV flux (blue solid line) and the observed-frame 0.5--2~keV flux (red solid line). 
We note that our sample is moderately complete from $f_{2-8 \text{ keV}}\approx 2\times 10^{-16}$ to $5\times10^{-15} \text{ erg s}^{-1}\text{ cm}^{-2}$ and $f_{0.5-2 \text{ keV}}\approx 5\times 10^{-17}$ to $2\times10^{-15} \text{ erg s}^{-1}\text{ cm}^{-2}$, with a steady increase to $100\%$  beyond these fluxes.
}  
 \label{fig:spec_comp_flux}
\end{figure}

In Figure~\ref{fig:spec_comp_flux}, we show the fraction of spectroscopically 
identified sources as a function of observed-frame 2--8~keV and 0.5--2~keV flux ($f_{2-8 \text{ keV}}$ and $f_{0.5-2 \text{ keV}}$, respectively). For the hard X-ray fluxes, we are highly complete ($>80\%$) above 
$f_{2-8 \text{ keV}} \approx 10^{-14} \text{ erg s}^{-1}\text{ cm}^{-2}$,
with the fraction falling to $\approx55$\% at fainter fluxes. The soft X-ray flux measurements are deeper, resulting in $>80\%$ completeness above $f_{0.5-2 \text{ keV}} \approx 3\times 10^{-15} \text{ erg s}^{-1}\text{ cm}^{-2}$, with a minimum fraction of $\approx 60\%$ below this. 
The dependence on the flux arises for two reasons: (1) sources that are less luminous in the X-ray tend to be fainter in the optical, and (2) the fraction of BLAGNs---which are much easier to identify spectroscopically---increases with X-ray luminosity (see Figure~\ref{fig:lumdist_classes} below).

%%%%%%%
% FIGURE 1
%%%%%%%

%%%%%%%%%%%%%%%%%%%%%%%%%%
\subsection{Spectral Classifications}
%%%%%%%%%%%%%%%%%%%%%%%%%%
We optically classified all of our DEIMOS spectra except for one, due to its poor signal to noise. 
Our spectral classes consist of broad absorption line quasars (BALQSOs), BLAGNs, type 2 AGNs, star formers (SFs), absorbers, or stars. For convenience, we combined the BALQSOs with the BLAGNs for our subsequent analysis. Sources are BLAGNs if they exhibit broad ($\text{FWHM}> 2000 \text{ km s}^{-1}$) emission lines in C III], C IV, Mg II, [O II], Si IV, or Ly$\alpha$ \citep[e.g.,][]{Barger2005,Barger2019}. Based on these criteria, we have 26 BLAGNs, including three BALQSOs.

We combined these spectral classifications with those available in the literature. Regarding the original L09 catalog, we treated objects labeled `D' in their source notes as stars. We grouped the \cite{Saez2015} categories of spiral galaxies (SpGs), star-forming galaxies (SFGs), and Lyman Break Galaxies (LBGs) into our star formers category, and their elliptical galaxies (EllGs) into our absorbers category. We used Gaussian profile fitting to estimate the FWHM of the spectral lines for their AGN classifications that were not already in our spectral classifications. This adds nine more BLAGNs, bringing the total number of BLAGNs to 35. 
We assigned the SF classification to all six sources identified by \cite{Umehata_2019}, and we adopted the classification of \cite{Newman2010A} for source ID 193. 

\begin{figure}[h!]
\centering
\includegraphics[width=\columnwidth]{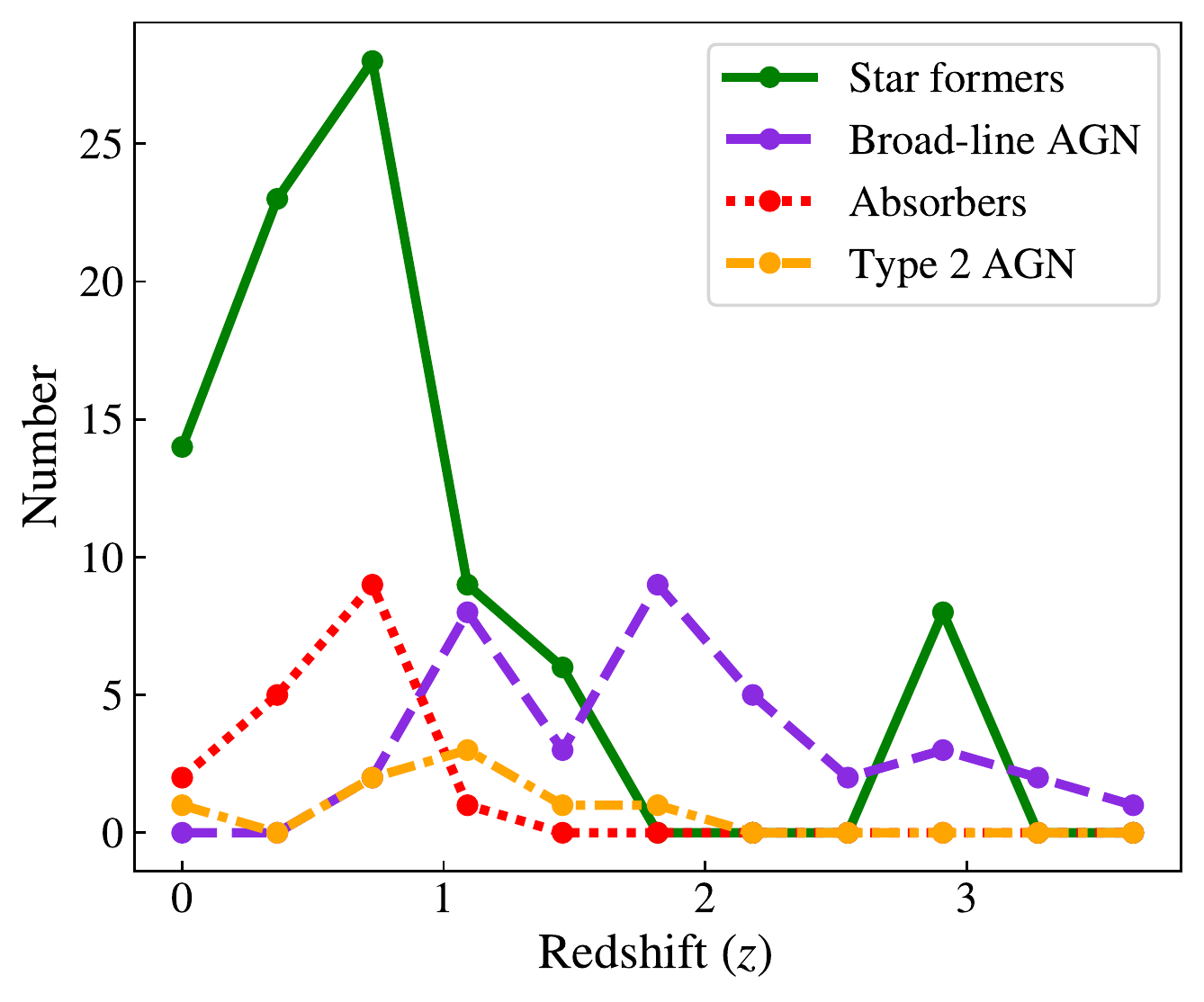}
\caption{Histograms show the distribution of optical spectral classification vs. redshift. The BLAGNs are most prominent at higher redshift ($z\gtrsim 2$), while sources undergoing enhanced rates of star formation dominate more recently ($0.25<z\lesssim1$). The spike of SFs near $z\sim3$ is due to the SSA22 protocluster on which the X-ray survey was centered.
}
\label{fig:ssa22_class}
\end{figure}

In total, we compile 165 optical spectral classifications for the X-ray sample, including 88 SFs, 35 BLAGNs, 17 absorbers, 17 stars, and 8 type 2 AGNs. We exclude all stars from further analysis. In Figure \ref{fig:ssa22_class}, we show the redshift distribution by classification, which reveals that SFs dominate the low-redshift population, while BLAGNs have a greater relative contribution at higher redshifts.

%%%%%%%
% FIGURE 2
%%%%%%%

In Table~\ref{tab:2}, we compile an updated catalog that incorporates all known X-ray fluxes, speczs, and optical spectral classifications for these 297 sources. We maintain the ordering of the source ID numbers (XIDs) given in L09 throughout this catalog.

%%%%%%%%%%%%%%%%%%%%%%%%%%
\subsection{Photometric Redshifts} \label{sec:photoz}
%%%%%%%%%%%%%%%%%%%%%%%%%%
In order to maximize the sample completeness and to avoid undercounting sources in the so-called optical ``redshift desert'' from $z\sim1.5$--$2$, we computed photometric redshifts (hereafter, photzs) for the X-ray sources. We utilized the \texttt{LePHARE} code \citep{Arnouts_2011} to fit a combination of template spectral energy distributions (SEDs) using all available observed broadband magnitudes, except for the UKIDSS $J$-band (which otherwise reduced the accuracy of our photzs). 
We used 10 galaxy and QSO templates from the SWIRE library \citep{Polletta_2007}, 10 composite templates \citep{Cristiani_2004, Netzer_2007, Rowan-Robinson_2008}, 8 galaxy SED templates from the PEGASE2 library \citep{Fioc_1997}, 5 built-in synthetic QSO templates, and 3 hybrids \citep{Salvato_2009}. We assumed the Calzetti dust law (e.g., \citealt{Calzetti_1997, Calzetti_2001, Calzetti_2000}), and we fit over a grid of color excess values from $E(B-V)=0$--0.35~mag for all sources.

To characterize our uncertainty, we computed two metrics derived from sources with both speczs and photzs: $\sigma_\text{diff}$ and $f_\text{outlier}$.  $\sigma_\text{diff}$ is the primary error estimate, as it is the standard deviation associated with the distribution of differences between spectroscopic and photometric values ($z_\text{diff}= z_\text{spec}-z_\text{phot}$). The outlier fraction $f_\text{outlier}$ is the fraction of spectroscopically confirmed sources whose photz satisfies $|
z_\text{diff}|/(1+z_\text{spec})>0.15$ (e.g., \citealt{Aird2015}). Overall, we find $\sigma_\text{diff}=0.63$ and $f_\text{outlier}=27\%$, suggesting that while not extremely precise, our photz estimates are sufficient for the goals of this work. We also computed photzs for our sample using \texttt{EAZY} \citep{BrammerEAZY2008ApJ...686.1503B} and found comparable results (but with slightly more scatter).

\begin{figure}[h!]
\centering
\includegraphics[width=\columnwidth]{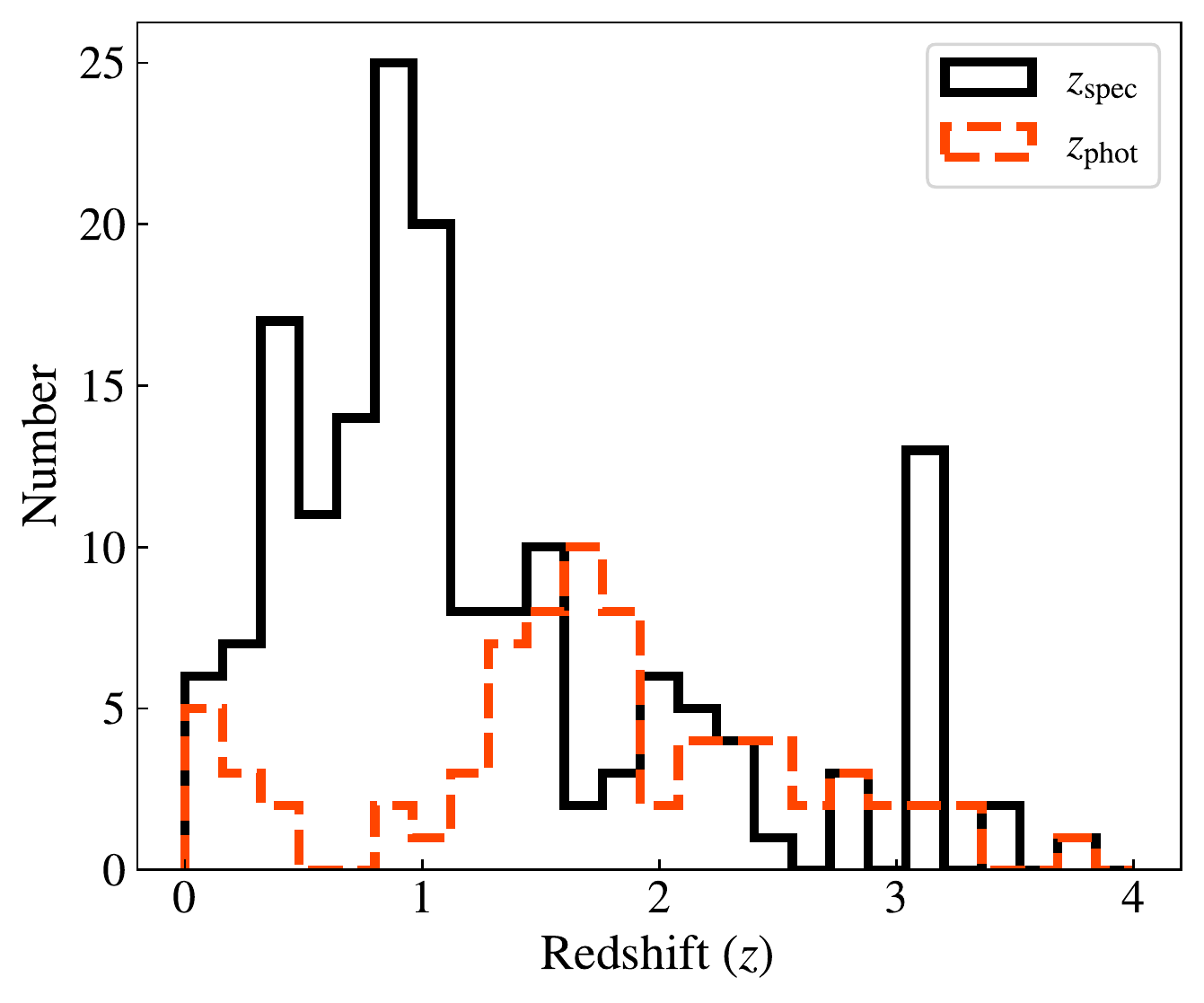}
\caption{Redshift distribution for the X-ray sources in our updated SSA22 catalog. Speczs are traced by the solid black line, while photzs from \texttt{LePHARE} are traced by the orange dashed line.
}
\label{fig:zspec_zphot_hist}
\end{figure}

We retained a total of 76 photzs, which brings our final extragalactic redshift count from 166 to 242 (these photzs are also presented in Table \ref{tab:2}). In Figure \ref{fig:zspec_zphot_hist}, we show the distributions of photzs and speczs, where we see a modest spike in photzs at low redshifts ($z_\text{phot}<0.3$). Each of the eight low-photz sources appears extremely faint in the optical and infrared (with several only having limits), which suggests their photzs may be more uncertain. However, nearly all of the probability distribution functions output by \texttt{LePHARE} for these sources show negligible probabilities for $z_\text{phot}\gtrsim 0.5$, so we adopt their computed photzs. Note that we attempted, unsuccessfully, to measure redshifts for six of these objects with DEIMOS.

%%%%%%%
% FIGURE 3
%%%%%%%

%%%%%%%%%%%%%%%%%%%%%%%%%%
\subsection{X-ray Luminosities}
\label{X-luminosity}
%%%%%%%%%%%%%%%%%%%%%%%%%%
Of the 166 extragalactic speczs, 11 have only 2--8~keV fluxes, 43 have only 0.5--2~keV fluxes, and 108 have both. To maximize the sensitivity of our analysis and to retain consistency in the energy range we are probing (rest-frame 2--8 keV), we divided our calculation of X-ray luminosity ($L_X$) into two definitions. At $z<3$ (where 150/166 galaxies lie), we derived our source luminosities using $f_{2-8 \text{ keV}}$. At $z\geq3$, we utilized deeper Chandra $f_{0.5-2 \text{ keV}}$ to calculate $L_X$. This technique ensures that we maximize our photometric completeness beyond $z=3$, while also achieving the best match between our low- and high-redshift data, as observed-frame 0.5--2~keV fluxes correspond exactly to rest-frame 2--8~keV fluxes at $z=3$ (e.g., \citealt{Yencho2009}). 

We assume the X-ray spectra in both redshift ranges take the form of a power law ($L_{\nu}\propto \nu^{\alpha}$; $\Gamma=1-\alpha$) with a photon index of $\Gamma =1.8$. Thus, our rest-frame 2--8 keV X-ray luminosities are calculated as: 
\begin{equation}
    L_{X}= 4\pi d_L^2\cdot
    \begin{cases}
        f_{2-8 \text{ keV}} \cdot (1+z)^{\Gamma-2} & \text{if } z<3 \\
        f_{0.5-2 \text{ keV}} \cdot ((1+z)/4)^{\Gamma-2}              & \text{if } z\geq3
    \end{cases}
    \label{Eq:L_X}
\end{equation}
where $d_L$ is the luminosity distance. Note that the K-correction is minimal for $z<1$, and vanishes at $z=3$. We restrict our analysis to sources obeying $L_X>10^{42} \text{ erg s}^{-1}$ to filter out stellar contamination (particularly X-ray binaries), making it highly probable that our X-ray sources are powered by an AGN (\citealt{10.1046/j.1365-8711.1998.01993.x}; \citealt{Moran1999}). We hereafter refer to any such objects as AGNs. We neglect absorption effects, as absorption corrections are relatively insignificant at 2--8 keV (\citealt{Cowie2003}).

\begin{figure}[h!]
\centering
\includegraphics[width=\columnwidth]{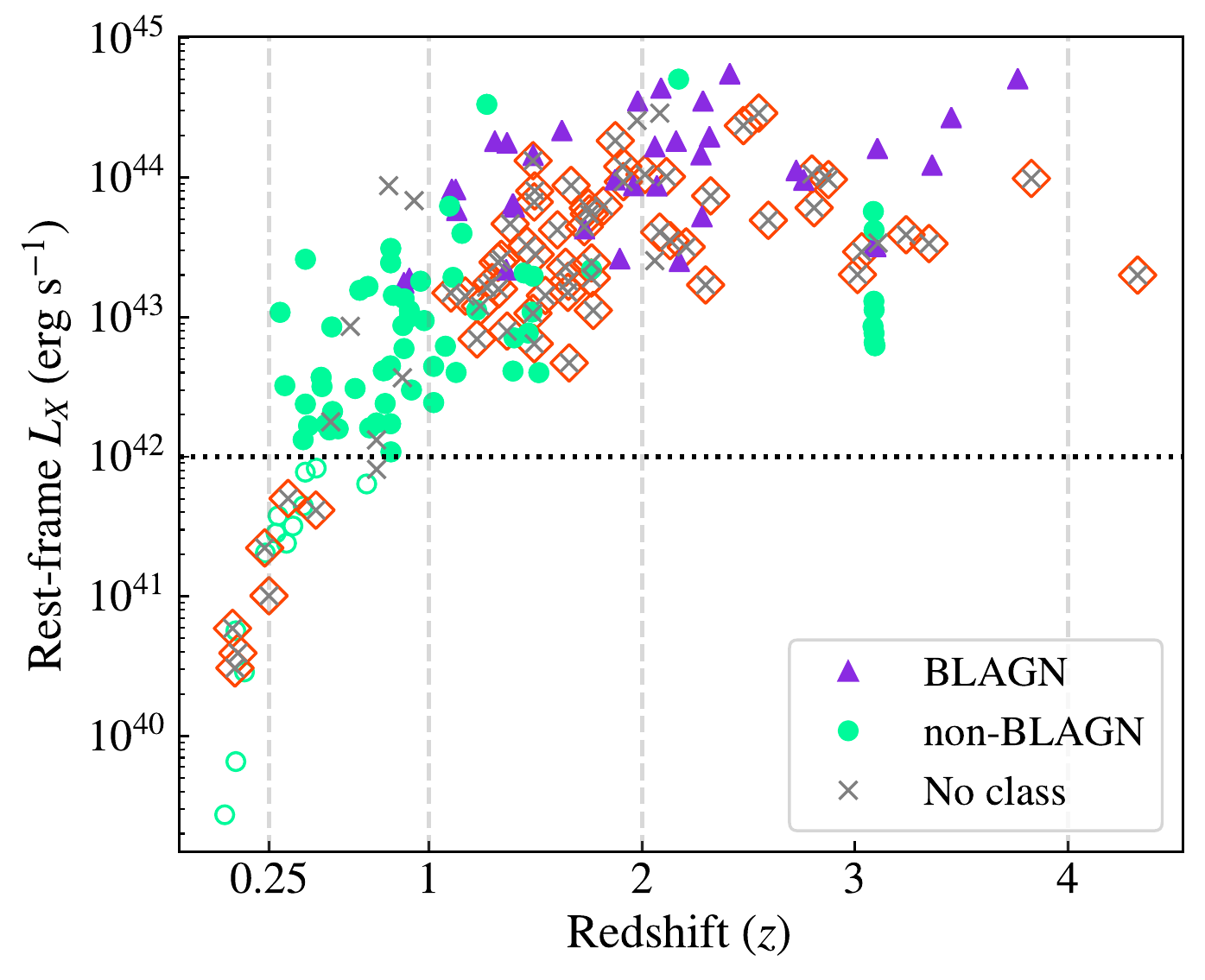}
\caption{The $L_X$ distribution of the full X-ray sample with rough classifications included. Violet triangles correspond to BLAGNs while filled green circles correspond to non-BLAGN (note that both populations are X-ray AGNs). Empty green circles represent non-BLAGNs that are also not AGNs and grey `x' marks trace sources that have no spectroscopic classification. Sources with photzs are highlighted by orange diamonds. The vertical dashed grey lines represent the redshift intervals over which we choose to construct our AGN XLFs, while the horizontal dotted black line corresponds to the AGN luminosity cut.}
\label{fig:lumdist_classes}
\end{figure}

To examine the luminosity distribution of AGNs across redshift, we split our AGN sample into BLAGN and non-BLAGN sub-samples. Specifically, the BLAGN sample consists of AGNs which have satisfied the FWHM criteria, while the non-BLAGN category is comprised of all other spectroscopically identified AGNs. As is well known, in
Figure~\ref{fig:lumdist_classes} we show how BLAGNs dominate at the highest luminosities, particularly for $L_X> 10^{43.5} \text{ erg s}^{-1}$ at $z\gtrsim1$. In the highest redshift interval ($z$=2--4), we find that 92\% of spectroscopically classified sources with $L_X>10^{44} \text{ erg s}^{-1}$ are BLAGNs. The mean overall BLAGN rest-frame 2--8~keV luminosity, across all redshifts, is $1.49\times 10^{44} \text{ erg s}^{-1}$ (compared to just $2.19\times 10^{43} \text{ erg s}^{-1}$ for our non-BLAGN sample). Conversely, non-BLAGNs are most common at lower X-ray luminosities near $L_X\approx10^{42}$--$10^{43} \text{ erg s}^{-1}$ (and lower redshifts), consistent with previous work (e.g., \citealt{Steffen2003}).

%%%%%%%
% FIGURE 4
%%%%%%%

%%%%%%%%%%%%%%%%%%%%%%%%%%
\section{The Redshift Evolution of the 2--8 keV X-ray Luminosity Function}
\label{sec:XLFs}
%%%%%%%%%%%%%%%%%%%%%%%%%%
We define the differential XLF as the number of AGNs per unit comoving volume, per unit logarithmic rest-frame 2--8 keV luminosity: $d\Phi(L_X, z)/(d\log L_X)$ (throughout this work, log refers to the base-10 logarithm). To calculate the XLF within a given redshift interval, we utilize a $1/V_a$ binning method \citep[see][]{1968ApJ...151..393S, 1977AJ.....82..861F, 1980ApJ...235..694A}, with equally sized luminosity bins $\Delta \log L_X$ within each redshift interval as
\begin{equation}
\frac{d \Phi(L_X,z)}{d \log L_X}\approx\Phi(L_X,z)=\frac{N/V_a}{\Delta \log L_X} \,,
\end{equation}
where $N$ is the number of sources with luminosities in the range $\Delta \log L_X$, spanning a unique comoving volume $V_a$ determined by the detector sensitivity.

In order to construct our final AGN XLF sample, we impose a number of cuts on the full data set, which we summarize in Table~\ref{tab:1}. Briefly, we select sources with redshifts (either speczs or photzs) in the range $0.25\leq z<4$ with measured rest-frame 2--8 keV fluxes (according to Equation~\ref{Eq:L_X}) that correspond to a luminosity consistent with AGNs ($L_X>10^{42} \text{ erg s}^{-1}$) and are $<10'$ off-beam. There are 155 objects that satisfy these criteria and therefore serve as the basis for all the XLFs in this work. We compare our results with the hard flux (2--10 keV) binned estimates given in Figure 7 of \cite{Aird2015} (hereafter, A15), which are available over our full redshift range. It should be noted that there is no source overlap between our samples and those of A15, though over 2800 of their X-ray sources are derived from five Chandra fields, including Bootes, Chandra Deep Field-South, Chandra Deep Field-North, Extended Groth Strip, and COSMOS. They supplemented these data with $\approx 150$ X-ray detections (most of which are AGNs) from wide-area surveys utilizing  ASCA (for the hard band) and ROSAT (for the soft band) to ensure coverage of the brightest X-ray sources. A15 also computed and included photzs (generated using \texttt{EAZY}). The A15 binned estimates we compare against were computed ignoring absorption effects and using the $N_\text{obs}/N_\text{mdl}$ method \citep{Miyaji_Hasinger_Schmidt2001,Aird2010}, which takes into account both the observed and predicted (per a model fit) number of sources in a given $L_X-z$ bin.

%%%%%%%%%%%%%%%%%%%%%%%%%%
\subsection{Comoving Volumes}
\begin{figure}[h!]
\centering
\includegraphics[width=\columnwidth]{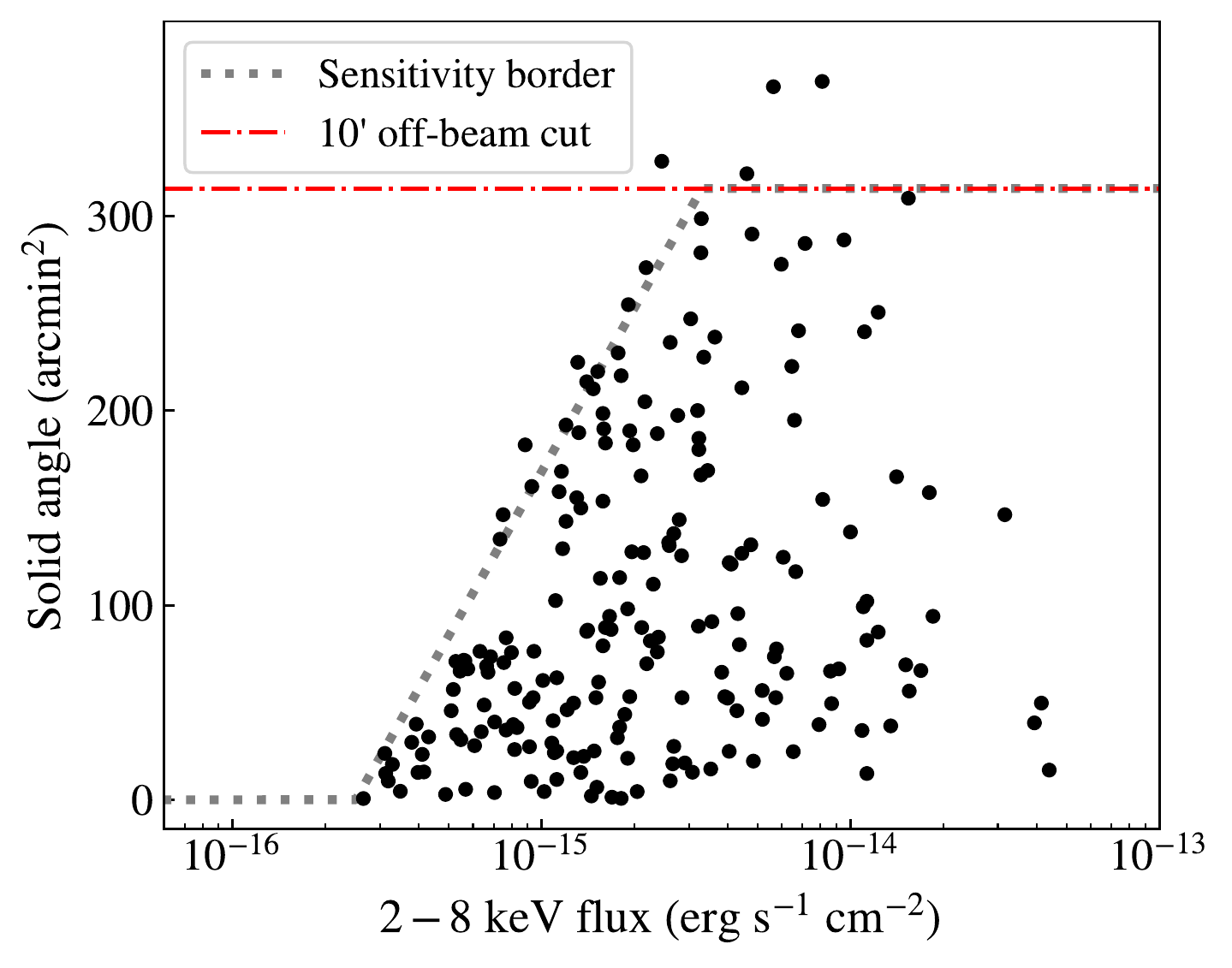}
\caption{The solid angle distribution for detected 2--8 keV fluxes. The grey dotted line represents the best-fit flux sensitivity border according to our log-linear fit of Equation \ref{Eq:Omega_eff}, while the red dashed line corresponds to the circular solid angle traced out by the $\theta=10'$ cutoff mark. As hard X-ray fluxes fall below $\sim 2\times 10^{-15}$~erg~s$^{-1}$ cm$^{-2}$, we see that detections are increasingly biased toward the beam center. 
}
\label{fig:hardfluxSA}
\end{figure}

%%%%%%%%%%%%%%%%%%%%%%%%%%
All the sources were observed with a single $16.9\times 16.9$ arcmin ACIS field of view (L09). We characterize the effect of detector sensitivity on our comoving volume calculation by fitting the apparent limiting boundary of the 2--8 keV flux as a function of the Chandra ACIS off-beam axis angle area subtended by each source's position in the field. We assume the functional form of this boundary to be a line in semi-log space $\Omega_\text{eff}(f) = af + b$, where $f$ is $\log (f_{2-8 \text{ keV}})$ and $a$ and $b$ are free parameters which we determine by least squares fitting (with units of arcmin$^2\text{ keV}^{-1}$ and arcmin$^2$, respectively). We cut out all sources with off-axis angles ($\theta$) exceeding $10'$ (corresponding to a $\pi 10^2$ arcmin$^2$ solid angle), as the X-ray coverage this far off-beam is not deep enough to construct an adequately complete sample. Doing so flattens our effective area curve at fluxes brighter than $f_{2-8 \text{ keV}} \approx 3\times 10^{-15} \text{ erg s}^{-1}$. Thus, the effective area as a function of of hard flux takes the form:
\begin{equation}
\Omega_\text{eff}(f) \text{ }[\text{arcmin}^2]=
\begin{cases}
0 & \text{if } f<f_\text{min} \\
a f + b & \text{if } f_\text{min} \leq f\leq f_\text{max} \\
\pi 10^2 & \text{otherwise}
\end{cases}
\label{Eq:Omega_eff}
\end{equation}
where $f_\text{min}$ is the minimum measured hard X-ray flux in the catalog and $f_\text{max}$ corresponds to the flux such that $a \cdot f_\text{max}+b=\pi 10^2$ (i.e., at the intersection of our sensitivity border line and our off-axis angle cut). Then, the effective area probed in each $L_X - z$ bin for our XLFs is the integrated average over the bin's $f_{2-8 \text{ keV}}$ range:
\begin{equation}
\Omega_\text{eff,bin} = \left [ \int^{f_\text{max,bin}}_{f_\text{min,bin}} \Omega_\text{eff}(f) df \right ](f_\text{max,bin}-f_\text{min,bin})^{-1} \,,
\end{equation}
where $f_\text{min,bin}$ and $f_\text{max,bin}$ are the minimum and maximum possible hard fluxes detected in a given bin. We substitute this $\Omega_\text{eff,bin}$ for the standard solid angle term in computing our comoving volumes, which are unique to each $L_X-z$ bin used to construct XLFs in this work. The best-fit flux boundary we use is depicted as the grey dotted line in Figure \ref{fig:hardfluxSA}, which is analogous to Figure 4b of L09. Note that we only consider the 2--8 keV flux boundary since only 20/155 sources in our final AGN XLF sample have X-ray luminosities calculated from their $0.5-2$ keV flux.

%%%%%%%
% FIGURE 5
%%%%%%%

%%%%%%%
% TABLE 1
%%%%%%%
\begin{deluxetable}{cc}
\tablenum{1}
\tablecaption{Final AGN XLF Sample Properties \label{tab:1}}
%\tablewidth{1.}
\tablehead{
\colhead{$N$} & \colhead{Cut}
}
\decimalcolnumbers
\decimals
\startdata
297 & None \\
226 & $0.25 \leq z < 4$ \\
171 & AND (($z<3$ \& $f_{2-8 \text{ keV}}$) \\
    & OR ($z\geq 3$ \& $f_{0.5-2 \text{ keV}}$)) \\
160 & AND $L_X > 10^{42} \text{ erg s}^{-1}$ \\
155 & AND $\theta<10'$ \\
\enddata
\tablecomments{Tabulation of data cuts and remaining number of sources ($N$) corresponding to the application of each subsequent cut.}
\end{deluxetable}

%%%%%%%%%%%%%%%%%%%%%%%%%%
\subsection{Incompleteness Correction}
%%%%%%%%%%%%%%%%%%%%%%%%%%

To account for sources without speczs or photzs, we apply an incompleteness correction. For each such source, we determine which subset of combinations of rest-frame $L_X$ and $z$ bins in our XLF that the source could occupy given its observed-frame $f_{2-8 \text{ kev}}$. We then increase the source counts of this subset of bins, weighted by the the number of known redshift sources already in each bin with an observed-frame $f_{2-8 \text{ kev}}$ similar to that of the source with unknown redshift. We rescale these added counts such that the total counts allocated to each $L_X-z$ bin sum up to 1. We repeat this process for each subsequent source without a known redshift. We calculate the fractional contributions from all sources with unknown redshifts before adding these contributions to the XLF data to avoid unknown source -- unknown source cross-correlation.

%%%%%%%%%%%%%%%%%%%%%%%%%%
\subsection{AGN X-ray Luminosity Functions}
\label{sec:AGN_XLF}
%%%%%%%%%%%%%%%%%%%%%%%%%%

\begin{figure*}
\centering
\includegraphics[width=\linewidth, scale=1]{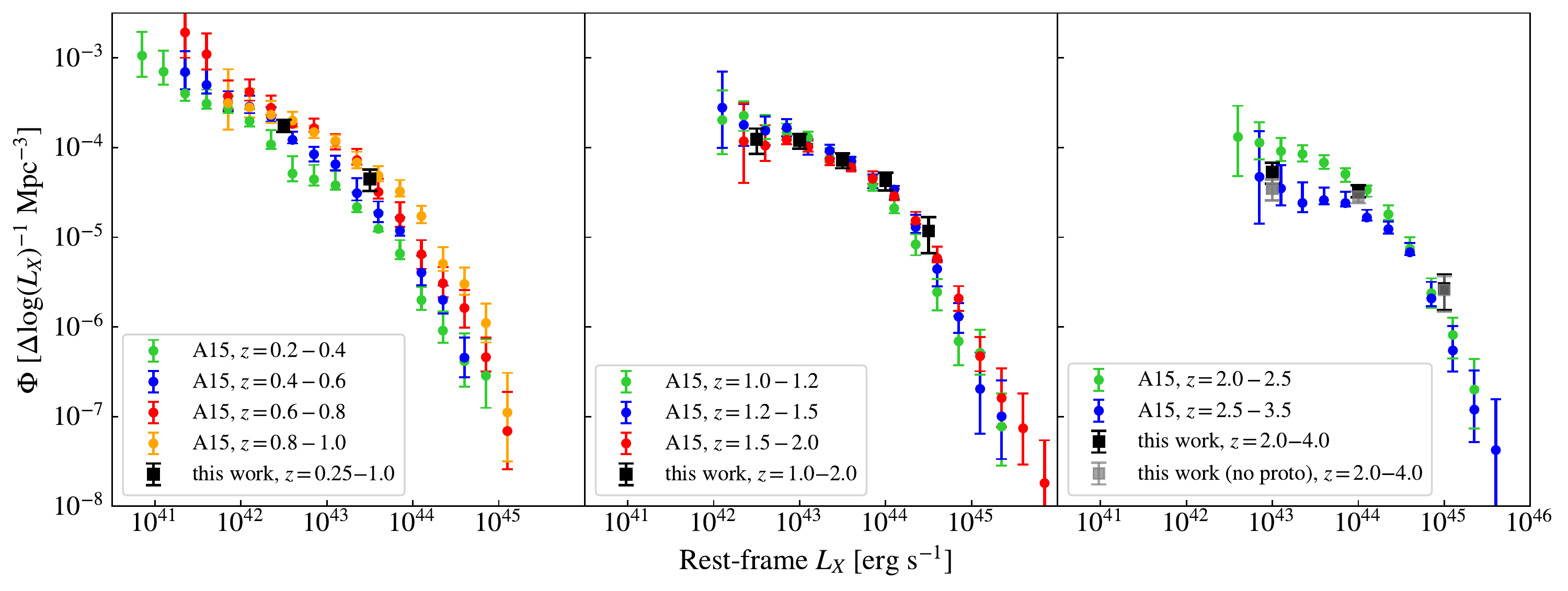}
\caption{The rest-frame 2--8 keV XLF for 155 AGNs in our SSA22 sample (black squares), split into three redshift intervals spanning $0.25\leq z < 4$. We overplot data from Figure 7 of A15 for overlapping redshift intervals (shown as green, blue, red, or orange). In the high redshift interval, we also plot the XLF excluding protocluster candidates (grey transparent squares). Vertical and horizontal axes share identical ranges. Our results are consistent with those of A15 to within $1\sigma$. }
\label{fig:ssa22_XLF_full}
\end{figure*}

We plot the incompleteness-corrected XLF of our full AGN sample, along with the results of A15, over the redshift intervals $z$=0.25--1, $z$=1--2 and $z$=2--4 in Figure \ref{fig:ssa22_XLF_full}. Black squares represent our bin centers ($\Phi$), of which there are two in the first interval, five in the second, and three in the final (each containing between 5 and 31 AGNs). Our error bars represent 1$\sigma$ Poissonian errors.

%%%%%%%
% FIGURE 6
%%%%%%%

We immediately note a steady decrease in the AGN number density and overall shift to higher 2--8 keV luminosities with increasing redshift. Next, although our full sample covers an on-sky area of only $\sim 0.08$ deg$^2$ while the Chandra surveys comprising A15's sample span $\sim 9$ deg$^2$ (and their large area surveys cover much more), our broadly-binned XLF closely reproduces their results. 

The SSA22 field is famous for the galactic protocluster at $z=3.09$, which may contain a significant fraction of the spectroscopically identified objects in our $z$=2--4 XLF interval. We investigated the effect of removing these protocluster candidates by excluding sources in our final AGN XLF sample with speczs in the redshift range $3.07<z<3.11$ and recomputing the XLF. In doing so, we repeat the full calculation, but we only display the high-redshift interval result (third panel of 
Figure~\ref{fig:ssa22_XLF_full}).

While we drop 13/47 (28\%) of our $z$=2--4 AGNs for this analysis, we find that our `no-proto' counts are largely comparable to our full sample counts. Without the protocluster candidates, the AGN densities fall, and thus the XLF becomes biased towards the higher-redshift ($z$=2.5--3.5) XLF of A15 at lower $L_X$ ($\lesssim 10^{43} \text{ erg s}^{-1}$). However, the AGN XLF appears to be increasingly insensitive to the presence of these sources at brighter X-ray luminosities and converges with our full-sample result near $L_X\approx 10^{45} \text{ erg s}^{-1}$.

%%%%%%%%%%%%%%%%%%%%%%%%%%
\subsection{The BLAGN and non-BLAGN XLFs}
\label{sec:classes_XLF}
\begin{figure*}
\centering
 \includegraphics[width=\textwidth, scale=1]{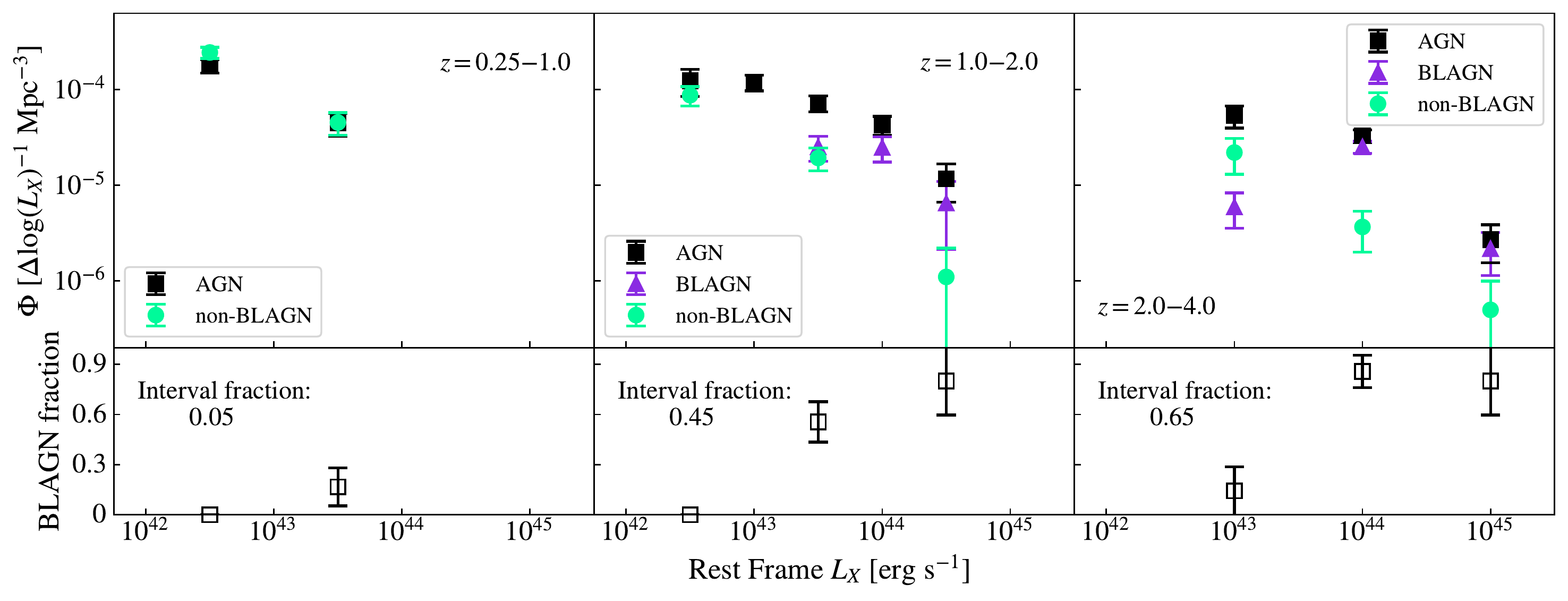}
\caption{(Top panel) The rest-frame 2--8 keV XLF for the 33 BLAGNs (violet triangles) and 61 non-BLAGNs (green circles) in our sample, constructed over the same redshift intervals and overplotted with our AGN XLF ($0.25\leq z<4$). Vertical and horizontal axes share identical ranges. (Bottom panel) The fraction of spectroscopically classified AGNs (94/155) in our AGN XLF sample that are BLAGNs, by redshift interval. Error bars are $1\sigma$ uncertainties derived from Monte Carlo simulations with $10^6$ Poissionian trials per bin.}
\label{fig:ssa22_XLF_BLAGN_nonblagn_BLAGNfraction}
\end{figure*}

%%%%%%%%%%%%%%%%%%%%%%%%%%
We now present the luminosity evolution of BLAGNs and all other spectroscopically confirmed AGNs. Our original sample of 35 BLAGNs reduces to 33 after applying our Table \ref{tab:1} cuts (note that all the BLAGNs satisfy our AGN luminosity cut). Since we are interested in the evolution of the non-BLAGNs, we derive our non-BLAGN sample from our AGN XLF sample of 155 objects. Of the 94 AGNs with secure classifications, 61 are non-BLAGNs. We construct the classified AGN XLFs within the previously defined $L_X-z$ bounds, with the exception of the $z$=0.25--1 interval, where we neglect BLAGNs altogether due to poor number statistics. In Figure~\ref{fig:ssa22_XLF_BLAGN_nonblagn_BLAGNfraction}, we show our BLAGN and non-BLAGN XLFs with our original AGN XLF overplotted, as well as the fraction of BLAGNs that constitute the spectroscopically classified AGN population within each interval.

%%%%%%%
% FIGURE 7
%%%%%%%

We note a few key properties within each interval. To begin, it is unsurprising that our non-BLAGN counts agree closely with our AGN XLF within the $z$=0.25--1 interval, as the majority of these AGNs are classified, and only two are BLAGNs. In our intermediate ($z$=1--2) and high redshift intervals ($z$=2--4), all sources are AGNs, and thus the relative contributions of BLAGNs and non-BLAGNs toward shaping our AGN XLF are clear. The non-BLAGN population constitutes the majority of the AGN number density at lower luminosities ($L_X < 10^{43.5} \text{ erg s}^{-1}$) with a sharp drop thereafter, while BLAGNs dominate the AGN XLF at the highest luminosities ($L_X > 10^{44} \text{ erg s}^{-1}$). Furthermore, the high-luminosity BLAGN XLF more closely matches the overall AGN XLF within the $z$=2--4 interval compared to the $z$=1--2 interval, suggesting that the AGN population is increasingly dominated by BLAGNs at higher redshift. These trends are matched by the evolution of the BLAGN fraction within each redshift interval, which increases with both luminosity and redshift.

%%%%%%%%%%%%%%%%%%%%%%%%%%
\section{Discussion} \label{sec:disc}
%%%%%%%%%%%%%%%%%%%%%%%%%%
One of the primary questions regarding the shape of the AGN XLF is whether the faint-end slope flattens at high redshift (e.g., \citealt{Hopkins2006}; \citealt{Aird2010}; A15). While we do not fit models to our AGN XLFs in this work, and we impose a conservative luminosity cut, which bars analysis of the faintest AGNs to increase the sample completeness, our results from Figure \ref{fig:ssa22_XLF_full} reproduce this flattening. As noted in previous work (e.g., A15), this is primarily driven by a changing mix of obscured non-BLAGNs and unobscured BLAGNs with increasing redshift. This is more clear in Figure~\ref{fig:ssa22_XLF_BLAGN_nonblagn_BLAGNfraction}, where we plot the BLAGN XLF along with the BLAGN fraction for each of our XLF intervals. BLAGNs become increasingly prevalent with redshift, but especially dominate (at the $>50\%$ level) the more luminous AGN population ($L_X \gtrsim 10^{43.5} \text{ erg~s}^{-1}$) beyond $z>1$. 

Despite our inability to probe the faint-end slope for the BLAGN and non-BLAGN XLFs separately (due to small number statistics), we still find distinct normalizations and break luminosities for each population. Our BLAGN XLF appears to have a lower normalization and higher break luminosity compared to the non-BLAGN population over the range $1\leq z<4$, showing agreement with A15's unabsorbed (BLAGN analog) and absorbed (non-BLAGN analog) AGN XLFs. This is evident in our intermediate- and high-redshift intervals, where we are able to plot both the BLAGN and the non-BLAGN XLFs. Overall, BLAGNs and non-BLAGNs both undergo luminosity evolution, tending toward higher X-ray luminosity with increasing redshift (i.e., from $z$=0.25 to $z$=4 for non-BLAGNs and from $z$=1 to $z$=4 for BLAGNs). 

While the density of non-BLAGNs at a given $L_X$ consistently decreases with redshift out to $z$=4, the density evolution of BLAGNs is slightly more complicated. That is, our high-redshift BLAGN XLF mostly exhibits lower densities than in the intermediate-redshift interval, but it shows a comparable (marginally enhanced) density for the bin centered on $L_X=10^{44} \text{ erg s}^{-1}$. These results are in agreement with A15, who found that absorbed AGNs exhibit stronger redshift evolution than their unabsorbed counterparts and make up a majority of the AGN population out to $z\sim 1$, before they drop off significantly at higher redshifts.

%%%%%%%%%%%%%%%%%%%%%%%%%%
\section{Summary} \label{conclusion}
%%%%%%%%%%%%%%%%%%%%%%%%%%

To briefly summarize this work:

\begin{enumerate}

 \item We updated and combined available spectroscopic redshifts and classifications for the \cite{Lehmer2009} SSA22 
 catalog of 297 X-ray sources. Our Keck/DEIMOS and LRIS spectra contributed 145 redshifts to the sample, with 41 novel 
 measurements, bringing the overall spectroscopic completeness to 183/297 or 62\%. 
 
\item We provided spectral classifications for most of our spectra and for others in the literature, bringing the number of classified 
sources to 165.
 
\item We constructed XLFs for 155 X-ray AGNs in the SSA22 field using speczs supplemented 
with photzs calculated from \texttt{LePHARE}. We found 
excellent agreement with the work of \cite{Aird2015} in each of our three redshift intervals: $z$=0.25-1, 
$z$=1--2, and $z$=2--4. Our results reproduce the flattening of the faint-end slope with redshift due to the changing contributions 
from BLAGNs and non-BLAGNs. Generally, the AGN XLF shifts to higher luminosities with increasing redshift.

\item We constructed the XLF for 33 BLAGNs and 61 non-BLAGNs separately, finding unique evolution for each 
population. Particularly, the density of BLAGNs remains lower than that of non-BLAGNs for lower luminosities 
($L_X \leq 10^{43.5} \text{ erg~s}^{-1}$), and higher for higher luminosities.
Furthermore, BLAGNs exhibit a brighter 
apparent break luminosity than non-BLAGNs from $z$=1 to $z$=4. Both populations shift to higher X-ray luminosities 
with redshift, but only non-BLAGNs exhibit substantial density evolution (from $z$=0.25--4), while the densities of BLAGNs
are more weakly evolving (from $z$=1--4). As a result of this, non-BLAGNs dominate the AGN population out to $z\sim 1$, 
while BLAGNs constitute the majority beyond $z\sim 2$, which is consistent with previous results.

\end{enumerate}

% %%%%%%
% TABLE 2
%%%%%%%
\begin{deluxetable*}{cccccccc}
\tablenum{2}
\tablecaption{Updated X-ray Source Catalog for SSA22 \label{tab:2}}
\tablewidth{0pt}
\tablehead{
\colhead{Source ID} & \multicolumn{2}{c}{X-ray Coordinates} &
\colhead{Spectroscopic} & \colhead{Photometric} & \colhead{0.5--2 keV Flux} & \colhead{2--8 keV Flux} & \colhead{Spectral} \\
\colhead{} & \colhead{RA J2000} & \colhead{DEC J2000} & \colhead{Redshift} &
\colhead{Redshift} & \colhead{} & \colhead{} & \colhead{Classification}
}
\decimalcolnumbers
\decimals
\startdata
1 & 22 16 51.96 & +00 18 49.00 & \nodata & 2.48 & 19.7 & 61.9 & \nodata \\
2 & 22 16 55.25 & +00 21 54.20 & \nodata & 2.65 & 16.7 & -48.2 & \nodata \\
3 & 22 16 56.32 & +00 16 57.70 & \nodata & 0.82 & 45.3 & -80.1 & \nodata \\
4 & 22 16 58.20 & +00 21 58.60 & 0.93 & 0.89 & 66.0 & 178 & \nodata \\
5 & 22 16 58.19 & +00 18 55.10 & \nodata & \nodata & 3.94 & 24.5 & \nodata \\
6 & 22 16 59.08 & +00 15 13.40 & 1.13 & 1.85 & 89.0 & 95.3 & BLAGN \\
7 & 22 17 00.33 & +00 19 55.20 & 2.277 & 2.21 & 26.3 & 46.2 & BALQSO \\
8 & 22 17 00.50 & +00 21 23.70 & 0.63 & 2.04 & 43.6 & 56.3 & \nodata \\
9 & 22 17 02.23 & +00 13 09.50 & \nodata & \nodata & -3.74 & 19.1 & \nodata \\
10 & 22 17 03.00 & +00 15 25.70 & 1.025 & 1.0 & -3.96 & -21.7 & \nodata \\
11 & 22 17 04.90 & +00 09 39.30 & 2.412 & 1.92 & 34.0 & 154 & BLAGN \\
12 & 22 17 05.41 & +00 15 14.00 & 3.765 & 3.25 & 39.4 & 65.9 & BLAGN \\
13 & 22 17 05.63 & +00 19 46.30 & \nodata & 1.6 & -3.76 & 30.4 & \nodata \\
14 & 22 17 05.82 & +00 22 27.70 & \nodata & \nodata & 3.03 & -32.0 & \nodata \\
15 & 22 17 05.83 & +00 22 24.70 & \nodata & 2.27 & -8.08 & -36.1 & \nodata \\
\nodata & \nodata & \nodata & \nodata & \nodata & \nodata & \nodata & \nodata \\
282 & 22 18 04.54 & +00 10 20.80 & 0.434 & 0.38 & 16.0 & 26.1 & Star Former \\
283 & 22 18 04.75 & +00 19 51.10 & \nodata & 1.91 & 17.2 & 44.5 & \nodata \\
284 & 22 18 05.04 & +00 14 02.00 & Star & \nodata & 3.88 & -14.5 & Star \\
285 & 22 18 05.36 & +00 12 49.10 & 2.175 & 2.13 & 6.39 & 8.85 & BLAGN \\
286 & 22 18 05.74 & +00 15 38.30 & \nodata & \nodata & 1.90 & -16.3 & \nodata \\
287 & 22 18 05.79 & +00 12 56.10 & \nodata & \nodata & 9.78 & 32.3 & \nodata \\
288 & 22 18 05.80 & +00 09 13.00 & 1.874 & 2.03 & 29.2 & 48.0 & BLAGN \\
289 & 22 18 05.98 & +00 15 06.70 & \nodata & 3.83 & 7.35 & 32.8 & \nodata \\
290 & 22 18 06.10 & +00 17 44.40 & Star & \nodata & -2.53 & 19.8 & Star \\
291 & 22 18 06.49 & +00 13 16.90 & 0.419 & \nodata & -3.00 & 13.2 & Star Former \\
292 & 22 18 06.86 & +00 14 18.30 & 0.864 & \nodata & 2.34 & -14.7 & Absorber \\
293 & 22 18 11.41 & +00 13 02.80 & \nodata & \nodata & 14.7 & 123 & \nodata \\
294 & 22 18 11.52 & +00 10 39.10 & 0.756 & \nodata & 3.54 & -26.7 & Star Former \\
295 & 22 18 12.23 & +00 15 01.40 & \nodata & 1.5 & -11.2 & 67.9 & \nodata \\
296 & 22 18 12.65 & +00 09 03.70 & 1.06 & 1.07 & 4.47 & -34.4 & \nodata \\
297 & 22 18 18.36 & +00 12 06.60 & \nodata & \nodata & -17.3 & 81.0 & \nodata \\ % table2
\enddata
\tablecomments{Updated X-ray/redshift/classification catalog for our SSA22 sources, originally presented in L09. Note that both columns (6) and (7) are in units of $10^{-16}$ erg s$^{-1}$ cm$^{-2}$, with negative fluxes indicating an upper limit. The full table, which includes references for each specz, is available in machine-readable form in the electronic version of this work. 
}
\end{deluxetable*}

%%%%%%%%%%%%%%%%%%%%%%%%%%
\begin{acknowledgements}
%%%%%%%%%%%%%%%%%%%%%%%%%%
We thank the anonymous referee for their careful review and comments which helped us to improve the manuscript.
We gratefully acknowledge support for this research from the trustees of the William F. Vilas Estate (B.T.R., A.J.T.),
a Wisconsin Space Grant Consortium Graduate and Professional Research Fellowship (A.J.T.), a Sigma Xi Grant 
in Aid of Research (A.J.T.), and a Kellett Mid-Career Award and a WARF Named 
Professorship from the University of Wisconsin-Madison
Office of the Vice Chancellor for Research and Graduate Education
with funding from the Wisconsin Alumni Research Foundation (A.J.B.).

The W.~M.~Keck Observatory is operated as 
a scientific partnership among the California Institute of Technology, the 
University of California, and NASA, and was made possible by the generous financial 
support of the W.~M.~Keck Foundation.

We wish to recognize and acknowledge the very significant 
cultural role and reverence that the summit of Maunakea has always 
had within the indigenous Hawaiian community. We are most fortunate 
to have the opportunity to conduct observations from this mountain.

This research has made use of the NASA/IPAC Extragalactic Database (NED),
which is operated by the Jet Propulsion Laboratory, California Institute of Technology,
under contract with the National Aeronautics and Space Administration.

\end{acknowledgements}

\facilities{Keck~I, Keck~II, CFHT, Subaru, Spitzer, UKIRT}

\software{\cite{Arnouts_2011}}

\bibliography{sample63}

\end{document}